%% file: main.tex
\documentclass[conference]{IEEEtran}
\usepackage{amsmath,graphicx,amssymb,amsfonts}
\usepackage{booktabs}
\usepackage{hyperref}
\usepackage{xcolor}
\usepackage{multirow}
\usepackage{array}
\usepackage{dsfont}
\usepackage{amssymb}
\usepackage{cite}
\usepackage{pifont}
\usepackage{amsthm}

\def\BibTeX{{\rm B\kern-.05em{\sc i\kern-.025em b}\kern-.08em
    T\kern-.1667em\lower.7ex\hbox{E}\kern-.125emX}}
\hypersetup{
    colorlinks=true,
    linkcolor=blue,
    filecolor=magenta,      
    urlcolor=cyan,
    pdfpagemode=FullScreen,
    }

\begin{document}

\title{Exploring Prediction Targets in Masked Pre-Training for Speech Foundation Models}

\author{
\IEEEauthorblockN{\begin{tabular}{c}Li-Wei Chen$^{\ddagger*}$, Takuya Higuchi$^{\dagger}$, He Bai$^{\dagger}$, Ahmed Hussen Abdelaziz$^{\dagger}$, Shinji Watanabe$^{\ddagger}$,\\ Alexander Rudnicky$^{\ddagger}$, Tatiana Likhomanenko$^{\dagger}$, Barry-John Theobald$^{\dagger}$, Zakaria Aldeneh$^{\dagger}$\end{tabular}}
\IEEEauthorblockA{\textit{$^{\ddagger}$Carnegie Mellon University} \qquad \textit{$^{\dagger}$ Apple}}}

\theoremstyle{remark}
\newtheorem*{remark}{Remark}
%
\maketitle
\renewcommand*{\thefootnote}{\fnsymbol{footnote}}
\footnotetext[1]{Work done during an internship at Apple.}
\renewcommand*{\thefootnote}{\arabic{footnote}}

\begin{abstract}
Speech foundation models, such as HuBERT and its variants, are pre-trained on large amounts of unlabeled speech data and then used for a range of downstream tasks. These models use a masked prediction objective, where the model learns to predict information about masked input segments from the unmasked context. The choice of prediction targets in this framework impacts their performance on downstream tasks. For instance, models pre-trained with targets that capture prosody learn representations suited for speaker-related tasks, while those pre-trained with targets that capture phonetics learn representations suited for content-related tasks. Moreover, prediction targets can differ in the level of detail they capture. Models pre-trained with targets that encode fine-grained acoustic features perform better on tasks like denoising, while those pre-trained with targets focused on higher-level abstractions are more effective for content-related tasks. Despite the importance of prediction targets, the design choices that affect them have not been thoroughly studied. This work explores the design choices and their impact on downstream task performance. Our results indicate that the commonly used design choices for HuBERT can be suboptimal. We propose approaches to create more informative prediction targets and demonstrate their effectiveness through improvements across various downstream tasks.
\end{abstract}
\section{Introduction}
\label{sec:intro}
Speech foundation models are trained on large amounts of unlabeled data using self-supervised learning (SSL). They can be used as pre-training  weights~\cite{Wav2Vec2,BEST-RQ}, or as feature extractors for lightweight prediction heads~\cite{SuPERB}. This paper focuses on the latter use case.
The success of SSL speech models depends on having a powerful encoder capable of producing features that are effective across a range of downstream tasks, including automatic speech recognition (ASR), speaker identification, and source separation. Consequently, numerous SSL approaches for learning encoders have been introduced (see~\cite{mohamed2022self} for a review).
Among these, a particularly successful family of models makes use of the \textit{masked prediction} objective, where the model is trained to reconstruct information randomly masked in the input from the unmasked context.
Notable examples in this family include HuBERT~\cite{HuBERT} and its derivatives~\cite{WavLM,UniSAT,contentvec,multiresHuBERT,chen2024robustspeechrepresentationlearning}, which we refer to collectively as HuBERT-based methods.

The choice of prediction targets is critical to the success of this paradigm.
Early works~\cite{TERA,Mockingjay,decoar} explored using low-level spectral features as prediction targets.
However, such targets are challenging to reconstruct due to their continuous and fine-grained nature~\cite{bai20223}. 
Consequently, later works~\cite{w2v-BERT,wav2vec-c,Wav2Vec2} explored methods for quantizing targets to abstract the fine-grained speech properties.
Wav2Vec 2.0~\cite{Wav2Vec2} designed a quantization module trained jointly with the masked prediction objective.
HuBERT~\cite{HuBERT} improved upon Wav2Vec 2.0 by replacing the quantization module with iterative clustering on learned features.
BEST-RQ~\cite{BEST-RQ} used random-projection quantizer to quantize speech signals to discrete labels.
Recent approaches~\cite{WavLM,contentvec,multiresHuBERT,UniSAT,chen2024robustspeechrepresentationlearning} built upon the iterative clustering framework of HuBERT with architectural changes and data augmentations.

The iterative clustering procedure used in HuBERT has been shown to improve representation learning of foundation models, and it is the primary focus of our study.
Iterative clustering involves key design decisions that affect the prediction targets.
These decisions, in turn, influence the performance of SSL features across various downstream tasks.
Note that HuBERT~\cite{HuBERT} original focused on content-based tasks, such as ASR, but widely adopted as a general foundation model~\cite{SuPERB,9747870,wang2022finetunedwav2vec20hubertbenchmark}. 
\textit{These observations motivate us to explore and adjust design decisions to support a wider range of downstream tasks.
}

We study how design decisions in the iterative clustering process of HuBERT-based methods affect the quality of the features for various downstream tasks. Specifically, we investigate design decisions that affect the prediction targets in two dimensions: 1) the content encoded and 2) the amount of information captured, which will be detailed in Section~\ref{sec:method}.
We analyze how variations in these two dimensions affect performance on downstream tasks.
We demonstrate that the widely used setup is suboptimal across the speech task.
We propose methods for enhancing the prediction targets, which attempt to improve the model's performance on phone recognition, speaker identification, and speech separation simultaneously.
Our systematic analysis on the design decisions provides useful guidance for research on masked prediction of speech.

\section{Method}
\label{sec:method}
Fig.~\ref{fig:explore} shows the commonly-used masked reconstruction speech pre-training framework~\cite{Wav2Vec2,HuBERT}.
The model uses convolutional layers to down-sample a given waveform into a sequence of dense representations.
Random masking is then applied, and transformer layers are trained to reconstruct the prediction targets of the masked portion.
To obtain the prediction targets, HuBERT-based approaches adopt an iterative clustering procedure that starts with prediction targets derived from clustered mel-frequency cepstral coefficients (MFCCs).
The procedure can be repeated multiple times by clustering representations from intermediate layers of the previous iteration and using them as targets for the next iteration.
Below, we introduce the design decisions that impact the content of the prediction targets and the amount of information they capture.
We further propose methods to enhance the prediction targets.

\input{figures/hubert_explore}
\subsection{Content of Prediction Targets}
\label{ssec:design_choices}
\subsubsection{Initial Target}
\label{ssec:initial_target}
The feature used to start the iterative procedure determines the prediction targets in the first iteration.
In prior work~\cite{HuBERT,WavLM,UniSAT}, MFCCs were used as initial features to cluster.
However, the extent to which the initial choice of features impacts the final performance of HuBERT remains unclear.
If the choice does have a significant impact, it encourages researchers to design starting features tailored to specific tasks.
Conversely, if the impact is minimal, it inspires further research to investigate the iterative process on demystifying its success. 
To this end, we study two additional settings for the initial features.
In the first setting, for the initial iteration, we train a model to predict the log Mel-spectrogram using the L1 loss and then cluster the resulting intermediate features.
In the second setting, we cluster features of a randomly initialized model to serve as the starting prediction targets.
The latter approach removes prior knowledge of speech from the training process, causing the training to be guided solely by the architecture of the neural network.
This approach is reminiscent of BEST-RQ~\cite{BEST-RQ}, where the prediction targets are derived from a random-projection quantizer.

\subsubsection{Layer to Cluster}
\label{ssec:layer_to_cluster}
Subsequent iterations in HuBERT-based methods cluster features from intermediate layers of the previous iteration model and use clusters as prediction targets.
Prior works~\cite{SuPERB,Layer-SSL,Layer-SSL2} have shown that different layers in pre-trained foundation models encode different aspects of speech.
For instance, higher layers of HuBERT were shown to encode more content information, while lower layers were shown to encode more speaker information.
As a result, the layer to cluster is expected to influence the information encoded in the prediction targets.
The HuBERT-based models selected the sixth layer for the second iteration and ninth layer for the third iteration for clustering.
However, these choices are not explicitly justified and tested on different tasks.
\subsubsection{\textbf{(Our Proposal)} Layer Multi-Target}
\label{ssec:layer_ensemble}
While downstream performance is sensitive to the choice of layer to cluster, conducting an exhaustive search across all layers to find the optimal clustering layer is computationally expensive.
To this end, we propose layer multi-target, predicting cluster IDs from all layers with a single foundation model.
We experimented with two methods.
In \textit{flat multi-target}, clusters from each layer are predicted independently using separate linear heads.
In contrast, \textit{conditional multi-target} predicts clusters of each layer conditioned on the ground-truth clusters of all higher layers.
For instance, when predicting the clusters of Layer 7, we provide the ground-truth clusters of Layer 9 and 11 to the prediction head.
This approach assumes higher layers contain refined information derived from the lower layers, which helps avoid redundant predictions by ensuring that each prediction head focuses on different aspects of the information.

\subsection{Information Granularity of Prediction Targets}
\subsubsection{Number of Clusters}
\label{ssec:n-clusters}
Prediction targets are obtained by clustering features via $k$-means, where each cluster represents a group of similar frames.
Having more clusters enables prediction targets to capture more fine-grained acoustic information.
Here, we examine downstream performances of models as the prediction targets capture progressively more detailed information.

\subsubsection{\textbf{(Our Proposal)} Residual Vector Quantization (RVQ) Tokens Prediction}
\label{ssec:rvq}
Adjusting the number of clusters allows us to explore how the resolution of prediction targets affects performance.
However, running $k$-means with a large number of clusters is computationally expensive.
Here, we explore an alternative approach to increase the information granularity in prediction targets.
Motivated by various studies~\cite{SoundStream,MQTTS,zhang2024speechtokenizer,shi24h_interspeech} showing that multiple quantizers capture fine detail in speech, we train foundation models that predict increasing levels of quantization tokens.
Specifically, we train a four-level RVQ-VAE~\cite{SoundStream} on log Mel-spectrogram and use increasing levels of learned RVQ tokens as prediction targets.
Focusing our analysis on iterative clustering, we combine RVQ-VAE with the clustering of HuBERT layers.
To this end, we modify the process of first-level quantization in RVQ-VAE, fixing the first-level code to the cluster indices.
Specifically, instead of using the closest code to the encoder output as done in the original RVQ-VAE, we use the code corresponding to the cluster index obtained from clustering layer nine of HuBERT.
We only modify the code-selection process, and the chosen code embeddings are still trained for reconstruction.
The remaining levels of quantizations follow the original RVQ-VAE training setup~\cite{SoundStream}.
Here we use \textit{conditional multi-target} detailed in Section~\ref{ssec:layer_ensemble} to predict multiple RVQ tokens.

\section{Experiment Setup}
We evaluate how design choices in HuBERT-based methods influence downstream performance.
To ensure a fair comparison, we isolate these design choices and fix HuBERT Base~\cite{HuBERT} as the model architecture.
Our models are pre-trained on the LibriSpeech dataset~\cite{LibriSpeech}, which contains $960$ hours of speech.
Unless otherwise noted, we perform clustering on the official HuBERT checkpoint\footnote{https://huggingface.co/facebook/hubert-base-ls960} (iteration two) to generate prediction targets for iteration three.
For $k$-means clustering, we use the \texttt{faiss} toolkit~\cite{faiss}.
We evaluate performance on SUPERB~\cite{SuPERB,superb-sg}, a widely used benchmark for speech foundation models.
To reduce computational burden, we focus on three representative tasks from the benchmark:

\paragraph{\textbf{(Content)} Phoneme Recognition (PR)} Phoneme recognition identifies the sequence of phonemes in target utterances.
We choose PR to represent content-based speech tasks, using Phone Error Rate (PER) as the evaluation metric.

\paragraph{\textbf{(Speaker)} Speaker Identification (SID)}
Speaker identification classifies utterances into a pre-defined set of  speakers.
We use SID to represent speaker-related tasks, using speaker classification accuracy (ACC) as the evaluation metric.

\paragraph{\textbf{(Acoustics)} Speech Separation (SS)}
Speech separation isolates target speech from background inferences.
We use SS to represent denoising tasks, using scale-invariant signal-to-distortion ratio improvement (SI-SDRi) as the metric. 

We follow the official setup of the SUPERB benchmark to train and evaluate all models. Thus, we refer the reader to the SUPERB paper~\cite{SuPERB} for additional details.

\section{Results}
\label{sec:results}
\subsection{Number of Iterations}
\input{tables/n_iters}
Before exploring the design decisions introduced in Section~\ref{sec:method}, we first investigate how the performance of HuBERT changes with each iteration on our target tasks.
Table~\ref{tab:n_iter} shows the results of different iterations.
We observe that the model converges\footnote{Convergence denotes no further improvement on any evaluated task.} by the third iteration for all tasks, with a substantial improvement from the second to third iteration.
As we want to compare the \textit{converged} performance of models, this result validates our choice to compare models in iteration three.
Note that the number of iterations required for convergence depends on the initial targets, as we will show in Section~\ref{ssec:exp-starting_features}.

\subsection{Content of Prediction Targets}
\subsubsection{Initial Targets}
\label{ssec:exp-starting_features}
\input{tables/start-point}
Table~\ref{tab:starting_point} presents a comparison of different initial targets as discussed in Section~\ref{ssec:initial_target}.
Our results show that the converged performance depends on the property of the initial targets.
For instance, MFCCs are widely used for ASR, and starting with MFCCs leads to the best PR performance.
log Mel-spectrogram contains more detailed spectral information than MFCCs, offering competitive SID and better SS performance compared to using MFCCs.
Starting from clusters of randomly initialized networks gives the best SS performance but leads to worse PR and SID performance.
We speculate that these random clusters, which are not designed to capture speech-relevant information, retain more acoustics-related information than log Mel-spectrogram.
These findings suggest that different initial targets reach different equilibriums after the iterative process, but there is no universally best initial targets for all the downstream tasks.
Additionally, having an initial target with prior knowledge of speech, such as MFCCs, effectively reduces the number of iterations required for convergence compared to random initialization.
Surprisingly, random initialization can achieve performance levels similar to MFCCs and log Mel-spectrograms with enough iterations.

\subsubsection{Layer to Cluster}
\label{ssec:exp-clustering_layers}
\input{tables/layers}
Table~\ref{tab:clustering_layer} presents the results when pre-training with clusters generated from different layers.
The results reaffirm that the choice of layer affects the downstream performance significantly.
Table~\ref{tab:clustering_layer} shows that deeper layers improve content-based performance compared to shallower layers.
However, this advantage does not hold for the other two tasks; for SS and SID, performance decreases when clustering uses deeper layers.
This result indicates that there is no single best layer for clustering across all speech tasks.
Although layer nine is typically chosen and performs well on PR, layer three outperforms layer nine on SS and SID tasks.

\subsubsection{\textbf{(Our Proposal)} Layer Multi-Target}
\label{ssec:exp-layer_ensemble}
Table~\ref{tab:clustering_layer} presents the performance of the layer multi-target proposed in Section~\ref{ssec:layer_ensemble}.
For fair comparison, we perform these methods on the $3^{rd},5^{th},7^{th},9^{th},11^{th}$ layers. \textit{Conditional multi-target} achieves better PR performance than clustering from any individual layer.
On the other hand, \textit{flat multi-target} gives the highest SID accuracy.
Both methods lead to competitive SS performance compared to the best performing individual layer (layer three).
These results indicate that layer multi-target is a good heuristic to bypass the laborious procedure of sweeping through individual layers.
More importantly, the results suggest the possibility of getting better performance by predicting more informative targets.

\subsection{Information Granularity of Prediction Targets}
\subsubsection{Number of Clusters}
\label{ssec:exp-n_clusters}
\input{tables/n_clusters}
Table~\ref{tab:n_clusters} shows how downstream performances vary with the number of clusters as discussed in Section~\ref{ssec:n-clusters}.
We observe a clear trend: increasing the number of clusters generally improves PR performance.
Notably, using more clusters results in a significant performance boost compared to the commonly used $500$ clusters, although it has little impact on other tasks.
Moreover, PR performance continues to improve up to $25000$ clusters, which is larger than the number of phoneme categories.
We attribute this improvement to the increased detail captured in coarticulation.
This finding shows that the overall performance benefits from predicting informative targets, in agreement with the results presented in Section~\ref{ssec:exp-layer_ensemble}.
Additionally, SID and SS performances fluctuated with the change in the number of clusters.

\subsubsection{\textbf{(Our Proposal)} RVQ Tokens Prediction}
\label{ssec:exp-rvq}
\input{tables/rvq_level}
As shown in Section~\ref{ssec:exp-layer_ensemble}, the proposed Layer Multi-Target approach improves performance, and as
discussed in Section~\ref{ssec:exp-n_clusters}, increasing the number of clusters also improves performance; we hypothesize
that predicting more information improves performance.
However, we also anticipated diminishing returns, as predicting excessive noise may not benefit content-based tasks.
To test this hypothesis, we experiment with the approach proposed in Section~\ref{ssec:rvq}.
The results are summarized in Table~\ref{tab:rvq}, where we increase the amount of information predicted by adding more quantizers.
This approach provides exponentially greater resolution compared to increasing the number of clusters.
We find that PR and SID performance peaks when predicting two levels of discrete tokens, but declines sharply after that point.
This result suggests there is an optimal amount of information for PR and SID tasks.
However, predicting additional levels of RVQ tokens consistently improves SS performance, which makes sense as the model needs to capture noise patterns to reconstruct higher-level RVQ tokens.

\input{figures/lw}
The SUPERB benchmark extracts representations from models with a weighted sum of transformer layers, which allows us to examine how layer contributions change as the levels of tokens increase.
Fig.~\ref{fig:lw} presents a visualization of this relationship for the three tasks.
For PR, adding third and fourth levels causes the large weights to occur at earlier layers.
This result indicates that fewer layers are used to process phonetic information, which could be one possible cause of the performance degradation of PR in Table~\ref{tab:rvq}.
For SS, more quantization levels increase the contribution of the last layer.
For SID, the best-performing model (+1RVQ) tends to have high weights concentrated on fewer layers.

\section{Conclusion}
This work investigated the relationship between the design decisions of HuBERT-based approaches and the downstream performance. We verified that the content of prediction targets noticeably affects downstream performance.
We showed that the widely used setup can be suboptimal by achieving better performance with more informative prediction targets.
Specifically, our proposed layer multi-target approach in Section~\ref{ssec:layer_ensemble} and RVQ token prediction in Section~\ref{ssec:rvq} provide better unified representation across phonetic, speaker, and acoustic properties of speech.

\vfill\pagebreak

\bibliographystyle{IEEEbib}
\bibliography{refs}

\end{document}

%% file: figures/hubert_explore.tex
\begin{figure}[tb]
    \centering
    \def\svgwidth{0.95\linewidth}
    \scriptsize
    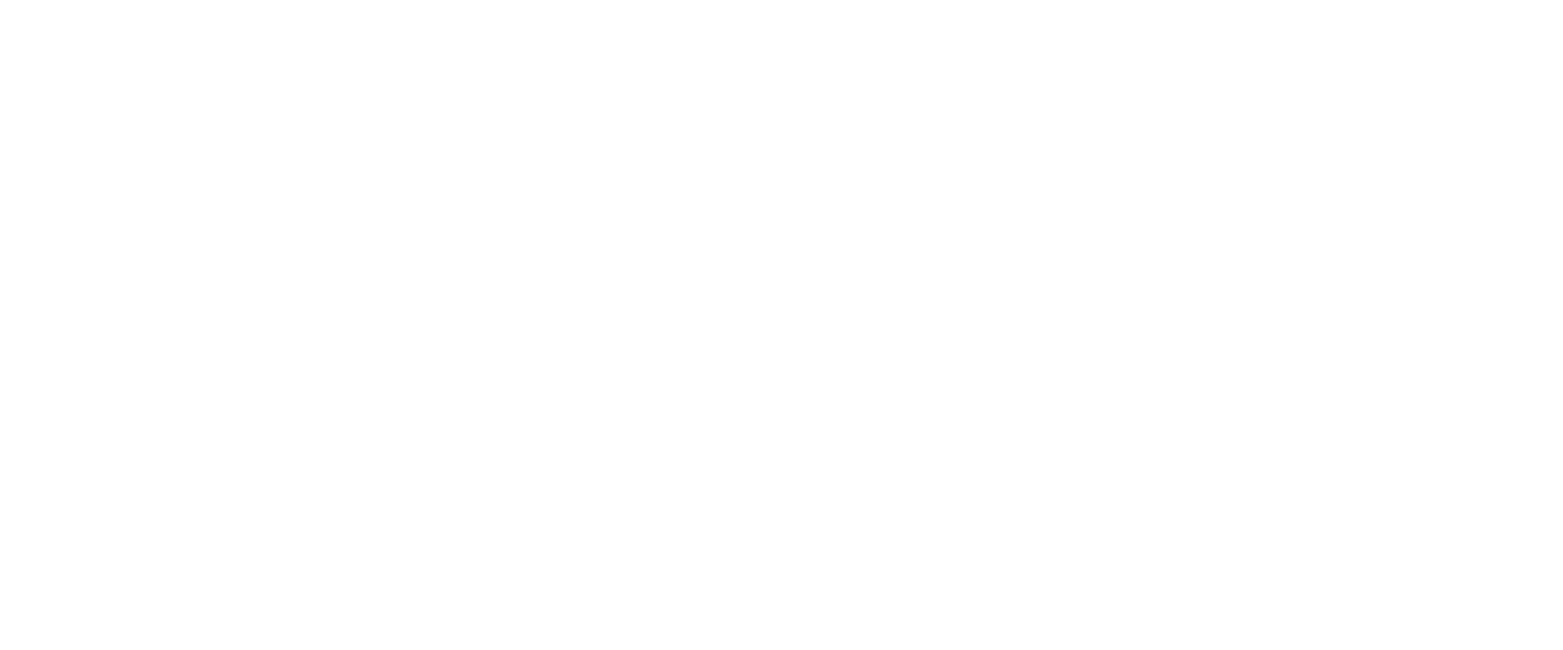
    \caption{Design decisions (marked in red) in the iterative clustering (HuBERT) procedure that affect prediction targets. 
    Detailed descriptions of these decisions are provided in Section~\ref{sec:method}.
}\label{fig:explore}
\end{figure}

%% file: figures_new/fig1_new_resize.pdf_tex
\begingroup%
  \makeatletter%
  \providecommand\color[2][]{%
    \errmessage{(Inkscape) Color is used for the text in Inkscape, but the package 'color.sty' is not loaded}%
    \renewcommand\color[2][]{}%
  }%
  \providecommand\transparent[1]{%
    \errmessage{(Inkscape) Transparency is used (non-zero) for the text in Inkscape, but the package 'transparent.sty' is not loaded}%
    \renewcommand\transparent[1]{}%
  }%
  \providecommand\rotatebox[2]{#2}%
  \newcommand*\fsize{\dimexpr\f@size pt\relax}%
  \newcommand*\lineheight[1]{\fontsize{\fsize}{#1\fsize}\selectfont}%
  \ifx\svgwidth\undefined%
    \setlength{\unitlength}{749.13999939bp}%
    \ifx\svgscale\undefined%
      \relax%
    \else%
      \setlength{\unitlength}{\unitlength * \real{\svgscale}}%
    \fi%
  \else%
    \setlength{\unitlength}{\svgwidth}%
  \fi%
  \global\let\svgwidth\undefined%
  \global\let\svgscale\undefined%
  \makeatother%
  \begin{picture}(1,0.41547189)%
    \lineheight{1}%
    \setlength\tabcolsep{0pt}%
    \put(0,0){\includegraphics[width=\unitlength,page=1]{figures_new/fig1_new_resize.pdf}}%
    \put(0.12102844,0.38671304){\makebox(0,0)[lt]{\lineheight{1.25}\smash{\begin{tabular}[t]{l}\textcolor{red}{Starting} \end{tabular}}}}%
    \put(0.06375286,0.35334145){\makebox(0,0)[lt]{\lineheight{1.25}\smash{\begin{tabular}[t]{l}\textcolor{red}{Prediction Targets}\end{tabular}}}}%
    \put(0,0){\includegraphics[width=\unitlength,page=2]{figures_new/fig1_new_resize.pdf}}%
    \put(0.05418549,0.1171725){\makebox(0,0)[lt]{\lineheight{1.25}\smash{\begin{tabular}[t]{l}Convolution Layers\end{tabular}}}}%
    \put(0,0){\includegraphics[width=\unitlength,page=3]{figures_new/fig1_new_resize.pdf}}%
    \put(0.0547193,0.24679073){\makebox(0,0)[lt]{\lineheight{1.25}\smash{\begin{tabular}[t]{l}Transformer Layers\end{tabular}}}}%
    \put(0,0){\includegraphics[width=\unitlength,page=4]{figures_new/fig1_new_resize.pdf}}%
    \put(0.10798809,0.00058882){\makebox(0,0)[lt]{\lineheight{1.25}\smash{\begin{tabular}[t]{l}Iteration 1\end{tabular}}}}%
    \put(0,0){\includegraphics[width=\unitlength,page=5]{figures_new/fig1_new_resize.pdf}}%
    \put(0.50673311,0.11696694){\makebox(0,0)[lt]{\lineheight{1.25}\smash{\begin{tabular}[t]{l}Convolution Layers\end{tabular}}}}%
    \put(0,0){\includegraphics[width=\unitlength,page=6]{figures_new/fig1_new_resize.pdf}}%
    \put(0.50726703,0.24658516){\makebox(0,0)[lt]{\lineheight{1.25}\smash{\begin{tabular}[t]{l}Transformer Layers\end{tabular}}}}%
    \put(0,0){\includegraphics[width=\unitlength,page=7]{figures_new/fig1_new_resize.pdf}}%
    \put(0.56053611,0.00038325){\makebox(0,0)[lt]{\lineheight{1.25}\smash{\begin{tabular}[t]{l}Iteration 2\end{tabular}}}}%
    \put(0.59205895,0.38650748){\makebox(0,0)[lt]{\lineheight{1.25}\smash{\begin{tabular}[t]{l}New \end{tabular}}}}%
    \put(0.51686603,0.35313587){\makebox(0,0)[lt]{\lineheight{1.25}\smash{\begin{tabular}[t]{l}Prediction Targets\end{tabular}}}}%
    \put(0,0){\includegraphics[width=\unitlength,page=8]{figures_new/fig1_new_resize.pdf}}%
    \put(0.32550671,0.21694611){\makebox(0,0)[lt]{\lineheight{1.25}\smash{\begin{tabular}[t]{l}\textcolor{red}{Number of} \end{tabular}}}}%
    \put(0.32550671,0.18357451){\makebox(0,0)[lt]{\lineheight{1.25}\smash{\begin{tabular}[t]{l}\textcolor{red}{Clusters}\end{tabular}}}}%
    \put(0,0){\includegraphics[width=\unitlength,page=9]{figures_new/fig1_new_resize.pdf}}%
    \put(0.32550671,0.13896109){\makebox(0,0)[lt]{\lineheight{1.25}\smash{\begin{tabular}[t]{l}\textcolor{red}{Layers to} \end{tabular}}}}%
    \put(0.32550671,0.10558949){\makebox(0,0)[lt]{\lineheight{1.25}\smash{\begin{tabular}[t]{l}\textcolor{red}{Cluster}\end{tabular}}}}%
    \put(0,0){\includegraphics[width=\unitlength,page=10]{figures_new/fig1_new_resize.pdf}}%
    \put(0.96586889,0.34509532){\makebox(0,0)[lt]{\lineheight{1.25}\smash{\begin{tabular}[t]{l}\textbf{…}\end{tabular}}}}%
    \put(0,0){\includegraphics[width=\unitlength,page=11]{figures_new/fig1_new_resize.pdf}}%
    \put(0.32864371,0.32651121){\makebox(0,0)[lt]{\lineheight{1.25}\smash{\begin{tabular}[t]{l}$k$-means \end{tabular}}}}%
    \put(0.78249732,0.32651121){\makebox(0,0)[lt]{\lineheight{1.25}\smash{\begin{tabular}[t]{l}$k$-means \end{tabular}}}}%
    \put(0.86029984,0.22503498){\makebox(0,0)[lt]{\lineheight{1.25}\smash{\begin{tabular}[t]{l}\textcolor{red}{Number of} \end{tabular}}}}%
    \put(0.86029984,0.19166339){\makebox(0,0)[lt]{\lineheight{1.25}\smash{\begin{tabular}[t]{l}\textcolor{red}{Iterations}\end{tabular}}}}%
    \put(0,0){\includegraphics[width=\unitlength,page=12]{figures_new/fig1_new_resize.pdf}}%
  \end{picture}%
\endgroup%

%% file: tables/n_iters.tex
\begin{table}[tb]
\centering
\caption{The number of iterations used in iterative clustering affects downstream performance. The performance converges on the third iteration.}
\label{tab:n_iter}
\begin{tabular}{l c c c}
\toprule
Model & PR/PER ↓ & SID/ACC(\%) ↑ & SS/SI-SDRi ↑ \\
\midrule
FBANK$^{\mathrm{\dagger}}$ & $82.01$ & $8.5$E$\text{-}4$ & $9.23$ \\
\midrule
Iter $1$ & $8.68$ & $72.57$ & $9.42$ \\
Iter $2^{\mathrm{*}}$ & $5.41$ & $81.42$ & $9.36$ \\
Iter $3$ & $\mathbf{4.72}$ & $\mathbf{81.82}$ & $\mathbf{9.59}$ \\
Iter $4$ & $4.80$ & $81.38$ & $\mathbf{9.59}$ \\
\bottomrule
\multicolumn{4}{l}{$^{\mathrm{*}}$Official HuBERT checkpoint.}\\
\multicolumn{4}{l}{$^{\mathrm{*,\dagger}}$Numbers are copied from SUPERB~\cite{SuPERB}.}
\end{tabular}
\end{table}

%% file: tables/start-point.tex
\begin{table}[tb]
\centering
\caption{Comparison of performance achieved with different initial targets. `Iter.' indicates the number of iterations required for convergence. `Mels' refers to log Mel-spectrogram. `Random' denotes clustering based on representations from randomly initialized networks.}
\label{tab:starting_point}
\begin{tabular}{l cc c c}
\toprule
Feature & Iter. & PR/PER ↓ & SID/ACC(\%) ↑ & SS/SI-SDRi ↑ \\
\midrule
MFCC$^{\mathrm{*}}$ & $3$ & $\mathbf{4.72}$ & $\mathbf{81.82}$ & $9.59$ \\
Mels & $3$ & $4.92$ & $81.80$ & $9.75$ \\
Random & $\mathbf{6}$ & $5.11$ & $79.99$ & $\mathbf{9.92}$ \\
\bottomrule
\multicolumn{4}{l}{$^{\mathrm{*}}$Commonly-used setup for HuBERT training.}
\end{tabular}
\end{table}

%% file: tables/layers.tex
\begin{table}[tb]
\centering
\caption{The impact of the layer used for generating prediction targets on downstream performance. Clustering is applied to features of the second iteration model, using $500$ clusters. 
`Cond.' refers to Conditional; see Section~\ref{ssec:layer_ensemble} for details on the proposed `Multi-target' methods.
}
\label{tab:clustering_layer}
\begin{tabular}{lc c c}
\toprule
Layers & PR/PER ↓ & SID/ACC(\%) ↑ & SS/SI-SDRi ↑ \\
\midrule
Layer $3$ & $5.99$ & $83.30$ & $9.77$ \\
Layer $5$ & $5.38$ & $82.46$ & $9.70$ \\
Layer $7$ & $5.01$ & $82.03$ & $9.54$ \\
Layer $9^{\mathrm{*}}$ & $4.72$ & $81.82$ & $9.59$ \\
Layer $11$ & $4.70$ & $82.20$ & $9.58$ \\
\midrule
Flat Multi-target & $4.72$ & $\mathbf{84.19}$ & $9.76$ \\
Cond. Multi-target & $\mathbf{4.49}$ & $82.37$ & $\mathbf{9.79}$ \\
\bottomrule
\multicolumn{4}{l}{$^{\mathrm{*}}$Commonly-used setup for HuBERT training.}

\end{tabular}
\end{table}

%% file: tables/n_clusters.tex
\begin{table}[tb]
\centering
\caption{The number of clusters used when generating prediction targets affects downstream performance. The clustering layer is fixed to the ninth layer.}
\label{tab:n_clusters}
\begin{tabular}{lc c c}
\toprule
\#Clusters & PR/PER ↓ & SID/ACC(\%) ↑ & SS/SI-SDRi ↑ \\
\midrule
$100$ & $4.78$ & $\mathbf{83.70}$ & $\mathbf{9.66}$ \\
$500^{\mathrm{*}}$ & $4.72$ & $81.82$ & $9.59$ \\
$2500$ & $4.47$ & $83.02$ & $9.59$ \\
$5000$ & $4.31$ & $81.41$ & $9.63$ \\
$10000$ & $4.16$ & $81.02$ & $9.64$ \\
$25000$ & $\mathbf{3.90}$ & $81.32$ & $9.64$ \\
\bottomrule
\multicolumn{4}{l}{$^{\mathrm{*}}$Commonly-used setup for HuBERT training.}
\end{tabular}
\end{table}

%% file: tables/rvq_level.tex
\begin{table}[tb]
\centering
\caption{Comparison of downstream performance across different quantization levels. +nRVQ indicates the use of additional Residual Vector Quantization (RVQ) levels alongside the original $k$-means tokens.}
\label{tab:rvq}
\begin{tabular}{l c c c}
\toprule
Tokens & PR/PER ↓ & SID/ACC(\%) ↑ & SS/SI-SDRi ↑ \\
\midrule
$k$-means$^{\mathrm{*}}$ & $4.72$ & $81.82$ & $9.59$ \\
\quad+$1$RVQ & $\mathbf{4.53}$ & $\mathbf{83.06}$ & $9.66$ \\
\quad+$2$RVQ & $5.80$ & $78.74$ & $9.84$ \\
\quad+$3$RVQ & $7.11$ & $76.32$ & $\mathbf{9.92}$ \\
\bottomrule
\multicolumn{4}{l}{$^{\mathrm{*}}$Commonly-used setup for HuBERT training.}
\end{tabular}
\end{table}

%% file: figures/lw.tex

\begin{figure}[tb]
    \centering
       %
\def\svgwidth{0.95\linewidth}
\tiny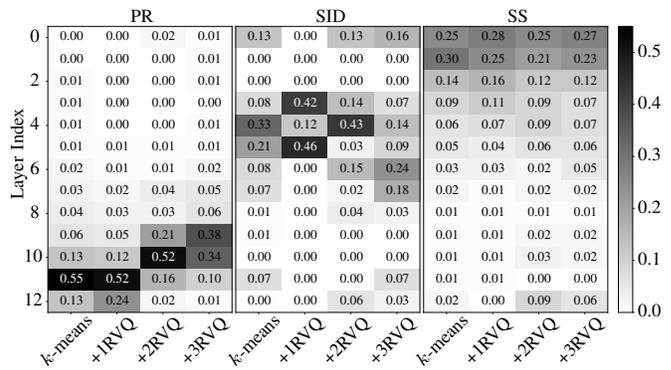
    \caption{Contribution of each transformer layer when predicting more informative targets. Evaluated on PR (left), SID (middle), and SS (right).
    Darker color means higher contribution. The $0^{th}$ layer refers to the input to the transformer.}
    \label{fig:lw}
\end{figure}

%% file: figures_new/heatmap_resize.pdf_tex
\begingroup%
  \makeatletter%
  \providecommand\color[2][]{%
    \errmessage{(Inkscape) Color is used for the text in Inkscape, but the package 'color.sty' is not loaded}%
    \renewcommand\color[2][]{}%
  }%
  \providecommand\transparent[1]{%
    \errmessage{(Inkscape) Transparency is used (non-zero) for the text in Inkscape, but the package 'transparent.sty' is not loaded}%
    \renewcommand\transparent[1]{}%
  }%
  \providecommand\rotatebox[2]{#2}%
  \newcommand*\fsize{\dimexpr\f@size pt\relax}%
  \newcommand*\lineheight[1]{\fontsize{\fsize}{#1\fsize}\selectfont}%
  \ifx\svgwidth\undefined%
    \setlength{\unitlength}{1226.05957031bp}%
    \ifx\svgscale\undefined%
      \relax%
    \else%
      \setlength{\unitlength}{\unitlength * \real{\svgscale}}%
    \fi%
  \else%
    \setlength{\unitlength}{\svgwidth}%
  \fi%
  \global\let\svgwidth\undefined%
  \global\let\svgscale\undefined%
  \makeatother%
  \begin{picture}(1,0.55973089)%
    \lineheight{1}%
    \setlength\tabcolsep{0pt}%
    \put(0,0){\includegraphics[width=\unitlength,page=1]{figures_new/heatmap_resize.pdf}}%
    \put(0.04545944,-0.00011491){\rotatebox{45}{\makebox(0,0)[lt]{\lineheight{1.25}\smash{\begin{tabular}[t]{l}{\scriptsize $k$-means}\end{tabular}}}}}%
    \put(0,0){\includegraphics[width=\unitlength,page=2]{figures_new/heatmap_resize.pdf}}%
    \put(0.12225531,0.00392223){\rotatebox{45}{\makebox(0,0)[lt]{\lineheight{1.25}\smash{\begin{tabular}[t]{l}{\scriptsize +1RVQ}\end{tabular}}}}}%
    \put(0,0){\includegraphics[width=\unitlength,page=3]{figures_new/heatmap_resize.pdf}}%
    \put(0.19501408,0.00392221){\rotatebox{45}{\makebox(0,0)[lt]{\lineheight{1.25}\smash{\begin{tabular}[t]{l}{\scriptsize +2RVQ}\end{tabular}}}}}%
    \put(0,0){\includegraphics[width=\unitlength,page=4]{figures_new/heatmap_resize.pdf}}%
    \put(0.26777284,0.00392223){\rotatebox{45}{\makebox(0,0)[lt]{\lineheight{1.25}\smash{\begin{tabular}[t]{l}{\scriptsize +3RVQ}\end{tabular}}}}}%
    \put(0,0){\includegraphics[width=\unitlength,page=5]{figures_new/heatmap_resize.pdf}}%
    \put(0.01863823,0.51457254){\makebox(0,0)[lt]{\lineheight{1.25}\smash{\begin{tabular}[t]{l}{\scriptsize 0}\end{tabular}}}}%
    \put(0,0){\includegraphics[width=\unitlength,page=6]{figures_new/heatmap_resize.pdf}}%
    \put(0.01863823,0.44500633){\makebox(0,0)[lt]{\lineheight{1.25}\smash{\begin{tabular}[t]{l}{\scriptsize 2}\end{tabular}}}}%
    \put(0,0){\includegraphics[width=\unitlength,page=7]{figures_new/heatmap_resize.pdf}}%
    \put(0.01863823,0.37544014){\makebox(0,0)[lt]{\lineheight{1.25}\smash{\begin{tabular}[t]{l}{\scriptsize 4}\end{tabular}}}}%
    \put(0,0){\includegraphics[width=\unitlength,page=8]{figures_new/heatmap_resize.pdf}}%
    \put(0.01863823,0.30587394){\makebox(0,0)[lt]{\lineheight{1.25}\smash{\begin{tabular}[t]{l}{\scriptsize 6}\end{tabular}}}}%
    \put(0,0){\includegraphics[width=\unitlength,page=9]{figures_new/heatmap_resize.pdf}}%
    \put(0.01863823,0.23630775){\makebox(0,0)[lt]{\lineheight{1.25}\smash{\begin{tabular}[t]{l}{\scriptsize 8}\end{tabular}}}}%
    \put(0,0){\includegraphics[width=\unitlength,page=10]{figures_new/heatmap_resize.pdf}}%
    \put(0.00885076,0.16674154){\makebox(0,0)[lt]{\lineheight{1.25}\smash{\begin{tabular}[t]{l}{\scriptsize 10}\end{tabular}}}}%
    \put(0,0){\includegraphics[width=\unitlength,page=11]{figures_new/heatmap_resize.pdf}}%
    \put(0.00885076,0.09717534){\makebox(0,0)[lt]{\lineheight{1.25}\smash{\begin{tabular}[t]{l}{\scriptsize 12}\end{tabular}}}}%
    \put(0.0074744,0.26511201){\rotatebox{90}{\makebox(0,0)[lt]{\lineheight{1.25}\smash{\begin{tabular}[t]{l}{\scriptsize Layer Index}\end{tabular}}}}}%
    \put(0,0){\includegraphics[width=\unitlength,page=12]{figures_new/heatmap_resize.pdf}}%
    \put(0.06562067,0.51666894){\makebox(0,0)[lt]{\lineheight{1.25}\smash{\begin{tabular}[t]{l}0.00\end{tabular}}}}%
    \put(0.13837943,0.51666894){\makebox(0,0)[lt]{\lineheight{1.25}\smash{\begin{tabular}[t]{l}0.00\end{tabular}}}}%
    \put(0.2111382,0.51666894){\makebox(0,0)[lt]{\lineheight{1.25}\smash{\begin{tabular}[t]{l}0.02\end{tabular}}}}%
    \put(0.28389696,0.51666894){\makebox(0,0)[lt]{\lineheight{1.25}\smash{\begin{tabular}[t]{l}0.01\end{tabular}}}}%
    \put(0.06562067,0.48188584){\makebox(0,0)[lt]{\lineheight{1.25}\smash{\begin{tabular}[t]{l}0.00\end{tabular}}}}%
    \put(0.13837943,0.48188584){\makebox(0,0)[lt]{\lineheight{1.25}\smash{\begin{tabular}[t]{l}0.00\end{tabular}}}}%
    \put(0.2111382,0.48188584){\makebox(0,0)[lt]{\lineheight{1.25}\smash{\begin{tabular}[t]{l}0.00\end{tabular}}}}%
    \put(0.28389696,0.48188584){\makebox(0,0)[lt]{\lineheight{1.25}\smash{\begin{tabular}[t]{l}0.01\end{tabular}}}}%
    \put(0.06562067,0.44710274){\makebox(0,0)[lt]{\lineheight{1.25}\smash{\begin{tabular}[t]{l}0.01\end{tabular}}}}%
    \put(0.13837943,0.44710274){\makebox(0,0)[lt]{\lineheight{1.25}\smash{\begin{tabular}[t]{l}0.00\end{tabular}}}}%
    \put(0.2111382,0.44710274){\makebox(0,0)[lt]{\lineheight{1.25}\smash{\begin{tabular}[t]{l}0.00\end{tabular}}}}%
    \put(0.28389696,0.44710274){\makebox(0,0)[lt]{\lineheight{1.25}\smash{\begin{tabular}[t]{l}0.01\end{tabular}}}}%
    \put(0.06562067,0.41231964){\makebox(0,0)[lt]{\lineheight{1.25}\smash{\begin{tabular}[t]{l}0.01\end{tabular}}}}%
    \put(0.13837943,0.41231964){\makebox(0,0)[lt]{\lineheight{1.25}\smash{\begin{tabular}[t]{l}0.00\end{tabular}}}}%
    \put(0.2111382,0.41231964){\makebox(0,0)[lt]{\lineheight{1.25}\smash{\begin{tabular}[t]{l}0.00\end{tabular}}}}%
    \put(0.28389696,0.41231964){\makebox(0,0)[lt]{\lineheight{1.25}\smash{\begin{tabular}[t]{l}0.00\end{tabular}}}}%
    \put(0.06562067,0.37753654){\makebox(0,0)[lt]{\lineheight{1.25}\smash{\begin{tabular}[t]{l}0.01\end{tabular}}}}%
    \put(0.13837943,0.37753654){\makebox(0,0)[lt]{\lineheight{1.25}\smash{\begin{tabular}[t]{l}0.00\end{tabular}}}}%
    \put(0.2111382,0.37753654){\makebox(0,0)[lt]{\lineheight{1.25}\smash{\begin{tabular}[t]{l}0.00\end{tabular}}}}%
    \put(0.28389696,0.37753654){\makebox(0,0)[lt]{\lineheight{1.25}\smash{\begin{tabular}[t]{l}0.01\end{tabular}}}}%
    \put(0.06562067,0.34275344){\makebox(0,0)[lt]{\lineheight{1.25}\smash{\begin{tabular}[t]{l}0.01\end{tabular}}}}%
    \put(0.13837943,0.34275344){\makebox(0,0)[lt]{\lineheight{1.25}\smash{\begin{tabular}[t]{l}0.01\end{tabular}}}}%
    \put(0.2111382,0.34275344){\makebox(0,0)[lt]{\lineheight{1.25}\smash{\begin{tabular}[t]{l}0.01\end{tabular}}}}%
    \put(0.28389696,0.34275344){\makebox(0,0)[lt]{\lineheight{1.25}\smash{\begin{tabular}[t]{l}0.01\end{tabular}}}}%
    \put(0.06562067,0.30797034){\makebox(0,0)[lt]{\lineheight{1.25}\smash{\begin{tabular}[t]{l}0.02\end{tabular}}}}%
    \put(0.13837943,0.30797034){\makebox(0,0)[lt]{\lineheight{1.25}\smash{\begin{tabular}[t]{l}0.01\end{tabular}}}}%
    \put(0.2111382,0.30797034){\makebox(0,0)[lt]{\lineheight{1.25}\smash{\begin{tabular}[t]{l}0.01\end{tabular}}}}%
    \put(0.28389696,0.30797034){\makebox(0,0)[lt]{\lineheight{1.25}\smash{\begin{tabular}[t]{l}0.02\end{tabular}}}}%
    \put(0.06562067,0.27318723){\makebox(0,0)[lt]{\lineheight{1.25}\smash{\begin{tabular}[t]{l}0.03\end{tabular}}}}%
    \put(0.13837943,0.27318723){\makebox(0,0)[lt]{\lineheight{1.25}\smash{\begin{tabular}[t]{l}0.02\end{tabular}}}}%
    \put(0.2111382,0.27318723){\makebox(0,0)[lt]{\lineheight{1.25}\smash{\begin{tabular}[t]{l}0.04\end{tabular}}}}%
    \put(0.28389696,0.27318723){\makebox(0,0)[lt]{\lineheight{1.25}\smash{\begin{tabular}[t]{l}0.05\end{tabular}}}}%
    \put(0.06562067,0.23840413){\makebox(0,0)[lt]{\lineheight{1.25}\smash{\begin{tabular}[t]{l}0.04\end{tabular}}}}%
    \put(0.13837943,0.23840413){\makebox(0,0)[lt]{\lineheight{1.25}\smash{\begin{tabular}[t]{l}0.03\end{tabular}}}}%
    \put(0.2111382,0.23840413){\makebox(0,0)[lt]{\lineheight{1.25}\smash{\begin{tabular}[t]{l}0.03\end{tabular}}}}%
    \put(0.28389696,0.23840413){\makebox(0,0)[lt]{\lineheight{1.25}\smash{\begin{tabular}[t]{l}0.06\end{tabular}}}}%
    \put(0.06562067,0.20362103){\makebox(0,0)[lt]{\lineheight{1.25}\smash{\begin{tabular}[t]{l}0.06\end{tabular}}}}%
    \put(0.13837943,0.20362103){\makebox(0,0)[lt]{\lineheight{1.25}\smash{\begin{tabular}[t]{l}0.05\end{tabular}}}}%
    \put(0.2111382,0.20362103){\makebox(0,0)[lt]{\lineheight{1.25}\smash{\begin{tabular}[t]{l}0.21\end{tabular}}}}%
    \put(0.28389696,0.20362103){\makebox(0,0)[lt]{\lineheight{1.25}\smash{\begin{tabular}[t]{l}0.38\end{tabular}}}}%
    \put(0.06562067,0.16883793){\makebox(0,0)[lt]{\lineheight{1.25}\smash{\begin{tabular}[t]{l}0.13\end{tabular}}}}%
    \put(0.13837943,0.16883793){\makebox(0,0)[lt]{\lineheight{1.25}\smash{\begin{tabular}[t]{l}0.12\end{tabular}}}}%
    \put(0.2111382,0.16883793){\makebox(0,0)[lt]{\lineheight{1.25}\smash{\begin{tabular}[t]{l}\textcolor{white}{0.52}\end{tabular}}}}%
    \put(0.28389696,0.16883793){\makebox(0,0)[lt]{\lineheight{1.25}\smash{\begin{tabular}[t]{l}0.34\end{tabular}}}}%
    \put(0.06562067,0.13405483){\makebox(0,0)[lt]{\lineheight{1.25}\smash{\begin{tabular}[t]{l}\textcolor{white}{0.55}\end{tabular}}}}%
    \put(0.13837943,0.13405483){\makebox(0,0)[lt]{\lineheight{1.25}\smash{\begin{tabular}[t]{l}\textcolor{white}{0.52}\end{tabular}}}}%
    \put(0.2111382,0.13405483){\makebox(0,0)[lt]{\lineheight{1.25}\smash{\begin{tabular}[t]{l}0.16\end{tabular}}}}%
    \put(0.28389696,0.13405483){\makebox(0,0)[lt]{\lineheight{1.25}\smash{\begin{tabular}[t]{l}0.10\end{tabular}}}}%
    \put(0.06562067,0.09927173){\makebox(0,0)[lt]{\lineheight{1.25}\smash{\begin{tabular}[t]{l}0.13\end{tabular}}}}%
    \put(0.13837943,0.09927173){\makebox(0,0)[lt]{\lineheight{1.25}\smash{\begin{tabular}[t]{l}0.24\end{tabular}}}}%
    \put(0.2111382,0.09927173){\makebox(0,0)[lt]{\lineheight{1.25}\smash{\begin{tabular}[t]{l}0.02\end{tabular}}}}%
    \put(0.28389696,0.09927173){\makebox(0,0)[lt]{\lineheight{1.25}\smash{\begin{tabular}[t]{l}0.01\end{tabular}}}}%
    \put(0.17754977,0.54626932){\makebox(0,0)[lt]{\lineheight{1.25}\smash{\begin{tabular}[t]{l}{\scriptsize PR}\end{tabular}}}}%
    \put(0,0){\includegraphics[width=\unitlength,page=13]{figures_new/heatmap_resize.pdf}}%
    \put(0.34231518,-0.00011488){\rotatebox{45}{\makebox(0,0)[lt]{\lineheight{1.25}\smash{\begin{tabular}[t]{l}{\scriptsize $k$-means}\end{tabular}}}}}%
    \put(0,0){\includegraphics[width=\unitlength,page=14]{figures_new/heatmap_resize.pdf}}%
    \put(0.41911104,0.00392226){\rotatebox{45}{\makebox(0,0)[lt]{\lineheight{1.25}\smash{\begin{tabular}[t]{l}{\scriptsize +1RVQ}\end{tabular}}}}}%
    \put(0,0){\includegraphics[width=\unitlength,page=15]{figures_new/heatmap_resize.pdf}}%
    \put(0.49186979,0.00392227){\rotatebox{45}{\makebox(0,0)[lt]{\lineheight{1.25}\smash{\begin{tabular}[t]{l}{\scriptsize +2RVQ}\end{tabular}}}}}%
    \put(0,0){\includegraphics[width=\unitlength,page=16]{figures_new/heatmap_resize.pdf}}%
    \put(0.56462859,0.00392223){\rotatebox{45}{\makebox(0,0)[lt]{\lineheight{1.25}\smash{\begin{tabular}[t]{l}{\scriptsize +3RVQ}\end{tabular}}}}}%
    \put(0,0){\includegraphics[width=\unitlength,page=17]{figures_new/heatmap_resize.pdf}}%
    \put(0.36247643,0.51666894){\makebox(0,0)[lt]{\lineheight{1.25}\smash{\begin{tabular}[t]{l}0.13\end{tabular}}}}%
    \put(0.43523519,0.51666894){\makebox(0,0)[lt]{\lineheight{1.25}\smash{\begin{tabular}[t]{l}0.00\end{tabular}}}}%
    \put(0.50799398,0.51666894){\makebox(0,0)[lt]{\lineheight{1.25}\smash{\begin{tabular}[t]{l}0.13\end{tabular}}}}%
    \put(0.58075274,0.51666894){\makebox(0,0)[lt]{\lineheight{1.25}\smash{\begin{tabular}[t]{l}0.16\end{tabular}}}}%
    \put(0.36247643,0.48188584){\makebox(0,0)[lt]{\lineheight{1.25}\smash{\begin{tabular}[t]{l}0.00\end{tabular}}}}%
    \put(0.43523519,0.48188584){\makebox(0,0)[lt]{\lineheight{1.25}\smash{\begin{tabular}[t]{l}0.00\end{tabular}}}}%
    \put(0.50799398,0.48188584){\makebox(0,0)[lt]{\lineheight{1.25}\smash{\begin{tabular}[t]{l}0.00\end{tabular}}}}%
    \put(0.58075274,0.48188584){\makebox(0,0)[lt]{\lineheight{1.25}\smash{\begin{tabular}[t]{l}0.00\end{tabular}}}}%
    \put(0.36247643,0.44710274){\makebox(0,0)[lt]{\lineheight{1.25}\smash{\begin{tabular}[t]{l}0.00\end{tabular}}}}%
    \put(0.43523519,0.44710274){\makebox(0,0)[lt]{\lineheight{1.25}\smash{\begin{tabular}[t]{l}0.00\end{tabular}}}}%
    \put(0.50799398,0.44710274){\makebox(0,0)[lt]{\lineheight{1.25}\smash{\begin{tabular}[t]{l}0.00\end{tabular}}}}%
    \put(0.58075274,0.44710274){\makebox(0,0)[lt]{\lineheight{1.25}\smash{\begin{tabular}[t]{l}0.00\end{tabular}}}}%
    \put(0.36247643,0.41231964){\makebox(0,0)[lt]{\lineheight{1.25}\smash{\begin{tabular}[t]{l}0.08\end{tabular}}}}%
    \put(0.43523519,0.41231964){\makebox(0,0)[lt]{\lineheight{1.25}\smash{\begin{tabular}[t]{l}\textcolor{white}{0.42}\end{tabular}}}}%
    \put(0.50799398,0.41231964){\makebox(0,0)[lt]{\lineheight{1.25}\smash{\begin{tabular}[t]{l}0.14\end{tabular}}}}%
    \put(0.58075274,0.41231964){\makebox(0,0)[lt]{\lineheight{1.25}\smash{\begin{tabular}[t]{l}0.07\end{tabular}}}}%
    \put(0.36247643,0.37753654){\makebox(0,0)[lt]{\lineheight{1.25}\smash{\begin{tabular}[t]{l}0.33\end{tabular}}}}%
    \put(0.43523519,0.37753654){\makebox(0,0)[lt]{\lineheight{1.25}\smash{\begin{tabular}[t]{l}0.12\end{tabular}}}}%
    \put(0.50799398,0.37753654){\makebox(0,0)[lt]{\lineheight{1.25}\smash{\begin{tabular}[t]{l}\textcolor{white}{0.43}\end{tabular}}}}%
    \put(0.58075274,0.37753654){\makebox(0,0)[lt]{\lineheight{1.25}\smash{\begin{tabular}[t]{l}0.14\end{tabular}}}}%
    \put(0.36247643,0.34275344){\makebox(0,0)[lt]{\lineheight{1.25}\smash{\begin{tabular}[t]{l}0.21\end{tabular}}}}%
    \put(0.43523519,0.34275344){\makebox(0,0)[lt]{\lineheight{1.25}\smash{\begin{tabular}[t]{l}\textcolor{white}{0.46}\end{tabular}}}}%
    \put(0.50799398,0.34275344){\makebox(0,0)[lt]{\lineheight{1.25}\smash{\begin{tabular}[t]{l}0.03\end{tabular}}}}%
    \put(0.58075274,0.34275344){\makebox(0,0)[lt]{\lineheight{1.25}\smash{\begin{tabular}[t]{l}0.09\end{tabular}}}}%
    \put(0.36247643,0.30797034){\makebox(0,0)[lt]{\lineheight{1.25}\smash{\begin{tabular}[t]{l}0.08\end{tabular}}}}%
    \put(0.43523519,0.30797034){\makebox(0,0)[lt]{\lineheight{1.25}\smash{\begin{tabular}[t]{l}0.00\end{tabular}}}}%
    \put(0.50799398,0.30797034){\makebox(0,0)[lt]{\lineheight{1.25}\smash{\begin{tabular}[t]{l}0.15\end{tabular}}}}%
    \put(0.58075274,0.30797034){\makebox(0,0)[lt]{\lineheight{1.25}\smash{\begin{tabular}[t]{l}0.24\end{tabular}}}}%
    \put(0.36247643,0.27318723){\makebox(0,0)[lt]{\lineheight{1.25}\smash{\begin{tabular}[t]{l}0.07\end{tabular}}}}%
    \put(0.43523519,0.27318723){\makebox(0,0)[lt]{\lineheight{1.25}\smash{\begin{tabular}[t]{l}0.00\end{tabular}}}}%
    \put(0.50799398,0.27318723){\makebox(0,0)[lt]{\lineheight{1.25}\smash{\begin{tabular}[t]{l}0.02\end{tabular}}}}%
    \put(0.58075274,0.27318723){\makebox(0,0)[lt]{\lineheight{1.25}\smash{\begin{tabular}[t]{l}0.18\end{tabular}}}}%
    \put(0.36247643,0.23840413){\makebox(0,0)[lt]{\lineheight{1.25}\smash{\begin{tabular}[t]{l}0.01\end{tabular}}}}%
    \put(0.43523519,0.23840413){\makebox(0,0)[lt]{\lineheight{1.25}\smash{\begin{tabular}[t]{l}0.00\end{tabular}}}}%
    \put(0.50799398,0.23840413){\makebox(0,0)[lt]{\lineheight{1.25}\smash{\begin{tabular}[t]{l}0.04\end{tabular}}}}%
    \put(0.58075274,0.23840413){\makebox(0,0)[lt]{\lineheight{1.25}\smash{\begin{tabular}[t]{l}0.03\end{tabular}}}}%
    \put(0.36247643,0.20362103){\makebox(0,0)[lt]{\lineheight{1.25}\smash{\begin{tabular}[t]{l}0.01\end{tabular}}}}%
    \put(0.43523519,0.20362103){\makebox(0,0)[lt]{\lineheight{1.25}\smash{\begin{tabular}[t]{l}0.00\end{tabular}}}}%
    \put(0.50799398,0.20362103){\makebox(0,0)[lt]{\lineheight{1.25}\smash{\begin{tabular}[t]{l}0.00\end{tabular}}}}%
    \put(0.58075274,0.20362103){\makebox(0,0)[lt]{\lineheight{1.25}\smash{\begin{tabular}[t]{l}0.00\end{tabular}}}}%
    \put(0.36247643,0.16883793){\makebox(0,0)[lt]{\lineheight{1.25}\smash{\begin{tabular}[t]{l}0.00\end{tabular}}}}%
    \put(0.43523519,0.16883793){\makebox(0,0)[lt]{\lineheight{1.25}\smash{\begin{tabular}[t]{l}0.00\end{tabular}}}}%
    \put(0.50799398,0.16883793){\makebox(0,0)[lt]{\lineheight{1.25}\smash{\begin{tabular}[t]{l}0.00\end{tabular}}}}%
    \put(0.58075274,0.16883793){\makebox(0,0)[lt]{\lineheight{1.25}\smash{\begin{tabular}[t]{l}0.00\end{tabular}}}}%
    \put(0.36247643,0.13405483){\makebox(0,0)[lt]{\lineheight{1.25}\smash{\begin{tabular}[t]{l}0.07\end{tabular}}}}%
    \put(0.43523519,0.13405483){\makebox(0,0)[lt]{\lineheight{1.25}\smash{\begin{tabular}[t]{l}0.00\end{tabular}}}}%
    \put(0.50799398,0.13405483){\makebox(0,0)[lt]{\lineheight{1.25}\smash{\begin{tabular}[t]{l}0.00\end{tabular}}}}%
    \put(0.58075274,0.13405483){\makebox(0,0)[lt]{\lineheight{1.25}\smash{\begin{tabular}[t]{l}0.07\end{tabular}}}}%
    \put(0.36247643,0.09927173){\makebox(0,0)[lt]{\lineheight{1.25}\smash{\begin{tabular}[t]{l}0.00\end{tabular}}}}%
    \put(0.43523519,0.09927173){\makebox(0,0)[lt]{\lineheight{1.25}\smash{\begin{tabular}[t]{l}0.00\end{tabular}}}}%
    \put(0.50799398,0.09927173){\makebox(0,0)[lt]{\lineheight{1.25}\smash{\begin{tabular}[t]{l}0.06\end{tabular}}}}%
    \put(0.58075274,0.09927173){\makebox(0,0)[lt]{\lineheight{1.25}\smash{\begin{tabular}[t]{l}0.03\end{tabular}}}}%
    \put(0.46984952,0.54626932){\makebox(0,0)[lt]{\lineheight{1.25}\smash{\begin{tabular}[t]{l}{\scriptsize SID}\end{tabular}}}}%
    \put(0,0){\includegraphics[width=\unitlength,page=18]{figures_new/heatmap_resize.pdf}}%
    \put(0.63917092,-0.00011489){\rotatebox{45}{\makebox(0,0)[lt]{\lineheight{1.25}\smash{\begin{tabular}[t]{l}{\scriptsize $k$-means}\end{tabular}}}}}%
    \put(0,0){\includegraphics[width=\unitlength,page=19]{figures_new/heatmap_resize.pdf}}%
    \put(0.7159668,0.00392225){\rotatebox{45}{\makebox(0,0)[lt]{\lineheight{1.25}\smash{\begin{tabular}[t]{l}{\scriptsize +1RVQ}\end{tabular}}}}}%
    \put(0,0){\includegraphics[width=\unitlength,page=20]{figures_new/heatmap_resize.pdf}}%
    \put(0.78872555,0.00392227){\rotatebox{45}{\makebox(0,0)[lt]{\lineheight{1.25}\smash{\begin{tabular}[t]{l}{\scriptsize +2RVQ}\end{tabular}}}}}%
    \put(0,0){\includegraphics[width=\unitlength,page=21]{figures_new/heatmap_resize.pdf}}%
    \put(0.86148435,0.00392223){\rotatebox{45}{\makebox(0,0)[lt]{\lineheight{1.25}\smash{\begin{tabular}[t]{l}{\scriptsize +3RVQ}\end{tabular}}}}}%
    \put(0,0){\includegraphics[width=\unitlength,page=22]{figures_new/heatmap_resize.pdf}}%
    \put(0.65933216,0.51666894){\makebox(0,0)[lt]{\lineheight{1.25}\smash{\begin{tabular}[t]{l}0.25\end{tabular}}}}%
    \put(0.73209093,0.51666894){\makebox(0,0)[lt]{\lineheight{1.25}\smash{\begin{tabular}[t]{l}0.28\end{tabular}}}}%
    \put(0.80484969,0.51666894){\makebox(0,0)[lt]{\lineheight{1.25}\smash{\begin{tabular}[t]{l}0.25\end{tabular}}}}%
    \put(0.87760845,0.51666894){\makebox(0,0)[lt]{\lineheight{1.25}\smash{\begin{tabular}[t]{l}0.27\end{tabular}}}}%
    \put(0.65933216,0.48188584){\makebox(0,0)[lt]{\lineheight{1.25}\smash{\begin{tabular}[t]{l}0.30\end{tabular}}}}%
    \put(0.73209093,0.48188584){\makebox(0,0)[lt]{\lineheight{1.25}\smash{\begin{tabular}[t]{l}0.25\end{tabular}}}}%
    \put(0.80484969,0.48188584){\makebox(0,0)[lt]{\lineheight{1.25}\smash{\begin{tabular}[t]{l}0.21\end{tabular}}}}%
    \put(0.87760845,0.48188584){\makebox(0,0)[lt]{\lineheight{1.25}\smash{\begin{tabular}[t]{l}0.23\end{tabular}}}}%
    \put(0.65933216,0.44710274){\makebox(0,0)[lt]{\lineheight{1.25}\smash{\begin{tabular}[t]{l}0.14\end{tabular}}}}%
    \put(0.73209093,0.44710274){\makebox(0,0)[lt]{\lineheight{1.25}\smash{\begin{tabular}[t]{l}0.16\end{tabular}}}}%
    \put(0.80484969,0.44710274){\makebox(0,0)[lt]{\lineheight{1.25}\smash{\begin{tabular}[t]{l}0.12\end{tabular}}}}%
    \put(0.87760845,0.44710274){\makebox(0,0)[lt]{\lineheight{1.25}\smash{\begin{tabular}[t]{l}0.12\end{tabular}}}}%
    \put(0.65933216,0.41231964){\makebox(0,0)[lt]{\lineheight{1.25}\smash{\begin{tabular}[t]{l}0.09\end{tabular}}}}%
    \put(0.7324478,0.41231964){\makebox(0,0)[lt]{\lineheight{1.25}\smash{\begin{tabular}[t]{l}0.11\end{tabular}}}}%
    \put(0.80484969,0.41231964){\makebox(0,0)[lt]{\lineheight{1.25}\smash{\begin{tabular}[t]{l}0.09\end{tabular}}}}%
    \put(0.87760845,0.41231964){\makebox(0,0)[lt]{\lineheight{1.25}\smash{\begin{tabular}[t]{l}0.07\end{tabular}}}}%
    \put(0.65933216,0.37753654){\makebox(0,0)[lt]{\lineheight{1.25}\smash{\begin{tabular}[t]{l}0.06\end{tabular}}}}%
    \put(0.73209093,0.37753654){\makebox(0,0)[lt]{\lineheight{1.25}\smash{\begin{tabular}[t]{l}0.07\end{tabular}}}}%
    \put(0.80484969,0.37753654){\makebox(0,0)[lt]{\lineheight{1.25}\smash{\begin{tabular}[t]{l}0.09\end{tabular}}}}%
    \put(0.87760845,0.37753654){\makebox(0,0)[lt]{\lineheight{1.25}\smash{\begin{tabular}[t]{l}0.07\end{tabular}}}}%
    \put(0.65933216,0.34275344){\makebox(0,0)[lt]{\lineheight{1.25}\smash{\begin{tabular}[t]{l}0.05\end{tabular}}}}%
    \put(0.73209093,0.34275344){\makebox(0,0)[lt]{\lineheight{1.25}\smash{\begin{tabular}[t]{l}0.04\end{tabular}}}}%
    \put(0.80484969,0.34275344){\makebox(0,0)[lt]{\lineheight{1.25}\smash{\begin{tabular}[t]{l}0.06\end{tabular}}}}%
    \put(0.87760845,0.34275344){\makebox(0,0)[lt]{\lineheight{1.25}\smash{\begin{tabular}[t]{l}0.06\end{tabular}}}}%
    \put(0.65933216,0.30797034){\makebox(0,0)[lt]{\lineheight{1.25}\smash{\begin{tabular}[t]{l}0.03\end{tabular}}}}%
    \put(0.73209093,0.30797034){\makebox(0,0)[lt]{\lineheight{1.25}\smash{\begin{tabular}[t]{l}0.03\end{tabular}}}}%
    \put(0.80484969,0.30797034){\makebox(0,0)[lt]{\lineheight{1.25}\smash{\begin{tabular}[t]{l}0.02\end{tabular}}}}%
    \put(0.87760845,0.30797034){\makebox(0,0)[lt]{\lineheight{1.25}\smash{\begin{tabular}[t]{l}0.05\end{tabular}}}}%
    \put(0.65933216,0.27318723){\makebox(0,0)[lt]{\lineheight{1.25}\smash{\begin{tabular}[t]{l}0.02\end{tabular}}}}%
    \put(0.73209093,0.27318723){\makebox(0,0)[lt]{\lineheight{1.25}\smash{\begin{tabular}[t]{l}0.01\end{tabular}}}}%
    \put(0.80484969,0.27318723){\makebox(0,0)[lt]{\lineheight{1.25}\smash{\begin{tabular}[t]{l}0.02\end{tabular}}}}%
    \put(0.87760845,0.27318723){\makebox(0,0)[lt]{\lineheight{1.25}\smash{\begin{tabular}[t]{l}0.02\end{tabular}}}}%
    \put(0.65933216,0.23840413){\makebox(0,0)[lt]{\lineheight{1.25}\smash{\begin{tabular}[t]{l}0.01\end{tabular}}}}%
    \put(0.73209093,0.23840413){\makebox(0,0)[lt]{\lineheight{1.25}\smash{\begin{tabular}[t]{l}0.01\end{tabular}}}}%
    \put(0.80484969,0.23840413){\makebox(0,0)[lt]{\lineheight{1.25}\smash{\begin{tabular}[t]{l}0.01\end{tabular}}}}%
    \put(0.87760845,0.23840413){\makebox(0,0)[lt]{\lineheight{1.25}\smash{\begin{tabular}[t]{l}0.01\end{tabular}}}}%
    \put(0.65933216,0.20362103){\makebox(0,0)[lt]{\lineheight{1.25}\smash{\begin{tabular}[t]{l}0.01\end{tabular}}}}%
    \put(0.73209093,0.20362103){\makebox(0,0)[lt]{\lineheight{1.25}\smash{\begin{tabular}[t]{l}0.01\end{tabular}}}}%
    \put(0.80484969,0.20362103){\makebox(0,0)[lt]{\lineheight{1.25}\smash{\begin{tabular}[t]{l}0.02\end{tabular}}}}%
    \put(0.87760845,0.20362103){\makebox(0,0)[lt]{\lineheight{1.25}\smash{\begin{tabular}[t]{l}0.02\end{tabular}}}}%
    \put(0.65933216,0.16883793){\makebox(0,0)[lt]{\lineheight{1.25}\smash{\begin{tabular}[t]{l}0.01\end{tabular}}}}%
    \put(0.73209093,0.16883793){\makebox(0,0)[lt]{\lineheight{1.25}\smash{\begin{tabular}[t]{l}0.01\end{tabular}}}}%
    \put(0.80484969,0.16883793){\makebox(0,0)[lt]{\lineheight{1.25}\smash{\begin{tabular}[t]{l}0.03\end{tabular}}}}%
    \put(0.87760845,0.16883793){\makebox(0,0)[lt]{\lineheight{1.25}\smash{\begin{tabular}[t]{l}0.02\end{tabular}}}}%
    \put(0.65933216,0.13405483){\makebox(0,0)[lt]{\lineheight{1.25}\smash{\begin{tabular}[t]{l}0.01\end{tabular}}}}%
    \put(0.73209093,0.13405483){\makebox(0,0)[lt]{\lineheight{1.25}\smash{\begin{tabular}[t]{l}0.01\end{tabular}}}}%
    \put(0.80484969,0.13405483){\makebox(0,0)[lt]{\lineheight{1.25}\smash{\begin{tabular}[t]{l}0.00\end{tabular}}}}%
    \put(0.87760845,0.13405483){\makebox(0,0)[lt]{\lineheight{1.25}\smash{\begin{tabular}[t]{l}0.00\end{tabular}}}}%
    \put(0.65933216,0.09927173){\makebox(0,0)[lt]{\lineheight{1.25}\smash{\begin{tabular}[t]{l}0.02\end{tabular}}}}%
    \put(0.73209093,0.09927173){\makebox(0,0)[lt]{\lineheight{1.25}\smash{\begin{tabular}[t]{l}0.00\end{tabular}}}}%
    \put(0.80484969,0.09927173){\makebox(0,0)[lt]{\lineheight{1.25}\smash{\begin{tabular}[t]{l}0.09\end{tabular}}}}%
    \put(0.87760845,0.09927173){\makebox(0,0)[lt]{\lineheight{1.25}\smash{\begin{tabular}[t]{l}0.06\end{tabular}}}}%
    \put(0.77256117,0.54626932){\makebox(0,0)[lt]{\lineheight{1.25}\smash{\begin{tabular}[t]{l}{\scriptsize SS}\end{tabular}}}}%
    \put(0,0){\includegraphics[width=\unitlength,page=23]{figures_new/heatmap_resize.pdf}}%
    \put(0.97865683,0.07978378){\makebox(0,0)[lt]{\lineheight{1.25}\smash{\begin{tabular}[t]{l}{\scriptsize 0.0}\end{tabular}}}}%
    \put(0,0){\includegraphics[width=\unitlength,page=24]{figures_new/heatmap_resize.pdf}}%
    \put(0.97865683,0.16199839){\makebox(0,0)[lt]{\lineheight{1.25}\smash{\begin{tabular}[t]{l}{\scriptsize 0.1}\end{tabular}}}}%
    \put(0,0){\includegraphics[width=\unitlength,page=25]{figures_new/heatmap_resize.pdf}}%
    \put(0.97865683,0.24421297){\makebox(0,0)[lt]{\lineheight{1.25}\smash{\begin{tabular}[t]{l}{\scriptsize 0.2}\end{tabular}}}}%
    \put(0,0){\includegraphics[width=\unitlength,page=26]{figures_new/heatmap_resize.pdf}}%
    \put(0.97865683,0.32642759){\makebox(0,0)[lt]{\lineheight{1.25}\smash{\begin{tabular}[t]{l}{\scriptsize 0.3}\end{tabular}}}}%
    \put(0,0){\includegraphics[width=\unitlength,page=27]{figures_new/heatmap_resize.pdf}}%
    \put(0.97865683,0.40864219){\makebox(0,0)[lt]{\lineheight{1.25}\smash{\begin{tabular}[t]{l}{\scriptsize 0.4}\end{tabular}}}}%
    \put(0,0){\includegraphics[width=\unitlength,page=28]{figures_new/heatmap_resize.pdf}}%
    \put(0.97865683,0.49085679){\makebox(0,0)[lt]{\lineheight{1.25}\smash{\begin{tabular}[t]{l}{\scriptsize 0.5}\end{tabular}}}}%
    \put(0,0){\includegraphics[width=\unitlength,page=29]{figures_new/heatmap_resize.pdf}}%
  \end{picture}%
\endgroup%